# Computocene: Notes from an Age of Observation

Simone Severini[1]

This piece plays with the idea of the *Computocene*: an era defined not merely by the ubiquity of computers, but by their deepening role in how we observe, interpret, and make sense of the world. Rather than emphasizing automation, speed, scale, or intelligence, computation is reframed as a mode of attention—filtering information, guiding inquiry, reframing questions, and shaping the very conditions under which knowledge emerges. I invite the reader to consider computers not simply as tools of calculation, but as epistemic instruments that participate in the formation of knowledge. This perspective reconfigures not only scientific practice but the epistemological foundations of understanding itself. The *Computocene* thus names a shift: from computation as calculation to computation as a form of attunement to the world. It is a speculative essay, offered without technical formality, and intended for a general, curious readership.

> *"The text is a machine conceived to elicit interpretations."*
> Umberto Eco [E1979]

## The Dawn of the Computocene

The *Computocene*, a term meant half in jest, half in earnest, describes an era not defined by computational ubiquity, but by a shift in how we observe and make sense of the world. Computation, I argue, may come to be seen less as a matter of speed and more as a matter of *attunement*: a way of noticing patterns and correlations that would otherwise remain inaccessible, and more *in tune* with how we think and relate to the world. I suggest that computers are evolving from instruments of calculation into epistemic machines, that is, devices that reshape the conditions under which knowledge emerges. This view deliberately avoids claims about mechanical intelligence. Even if the thesis proves false, imprecise, or unclear, it may still serve as an invitation to think differently about the epistemological role of computers and, conversely, their impact on our own cognition and our role in knowledge production. Rather than replacing human thought, they may redistribute it by assisting with a kind of *cognitive heavy lifting*. They open new domains of observation and interpretation, and with them, new modes of understanding. Thus, the Computocene is not about artificial intelligence *per se*, but about a reconfiguration of human attention in partnership with computational systems.

While this cognitive partnership evolves, the future of computation will increasingly rely on specialized hardware architectures, moving beyond the traditional paradigms. Moore's Law will become secondary, as performance gains will no longer be primarily driven by transistor density improvements. Instead, the focus will shift to algorithmic innovation and the intricate co-design of hardware and software systems, where computational solutions are tailored to specific tasks and domains. For instance, edge computing will play a pivotal role in this evolution, pushing processing closer to data sources and enabling more efficient, localized computational patterns. The related architectural diversification represents not just a technical necessity but a fundamental shift in how we approach computation, creating a more heterogeneous and distributed computational landscape that better aligns with the varied demands of modern cognitive tasks, either performed by humans or by robots. In all of this, physics-based computation will likely emerge as a vital component, offering superior performance for specific types of problems (see, for example, quantum computation).

---

[1] s.severini@ucl.ac.uk

With these ideas about new computational horizons in mind, we must remain clear-eyed about the current limitations of artificial intelligence systems. The gap between imaginary possibilities and practical realities remains significant, particularly in domains requiring even basic reasoning and a minimum of creative problem-solving. A telling example illustrates this disconnect.

*A negative preamble.* On April 1$^{st}$ 2025, mathematician Timothy Gowers posted a tongue-in-cheek message on the web. He claimed that after multiple failed attempts, he had finally gotten the AI model Grok to solve a challenging math problem: the "well-known Dubnový Blázen problem in graph theory," which he said he'd been working on for over a year [G2025]. He asked, half-seriously, "How long till it's better than human mathematicians across the board?" The joke was clear to those in the know, because "Dubnový Blázen" is Czech for "April Fool." The problem didn't exist.

A few days later, people discovered something more troubling: when searching for this fictional problem, AI-generated content began confidently describing it as real, claiming that Grok had indeed solved it after multiple attempts. What began as a prank quickly turned into a case study in misinformation. A fictional problem, invented as satire, had been absorbed into the *infosphere* (à la Floridi [F2014]) of AI-generated knowledge (or, simply, *knowledge*, given the indistinguishability). As *Private Eye* wryly concluded [P2025], "It's wonderful watching the well of knowledge being poisoned in real time." Yet the Computocene isn't simply a story of digital mirages. Consider, by contrast, a development that has already demonstrated how computers are transforming science.

*A positive preamble.* For half a century, determining the structure of proteins was one of the slowest (computationally speaking) parts of doing biology. The problem was well-defined: take a chain of tiny components and figure out how it folds in 3D space. The solutions were too complex to obtain and often elusive. Then came AlphaFold. In 2020, a team at DeepMind announced that their AI system could predict the three-dimensional shapes of proteins with startling accuracy [A2020].

By 2021, they released predictions for nearly every known protein, with a database covering over 200 million structures. It wasn't just a technical feat. It was a huge shift in tempo. Biologists started using it overnight. Structures that once took months to decode now took minutes (interestingly enough, walking around the swamps of computational complexity). It helped with enzyme design, drug targeting, and even filled in missing pieces of the nuclear pore complex, one of the cell's most intricate machines. In 2024, the Nobel Committee took notice [N2024]. Demis Hassabis and John Jumper, the minds behind AlphaFold, were awarded the Nobel Prize in Chemistry. Not for discovering a new molecule, but for building a machine that changed how molecules are discovered. Thus, this is a story beyond proteins. It's about what happens when scientific intuition can be obtained differently, when discovery becomes something that can be delegated, at least in part, to a machine.

These contrasting cases, one highlighting the potential for epistemic corruption, the other demonstrating unprecedented scientific breakthrough, exemplify the dual nature of our computational turn. They raise fundamental questions about how we validate and create knowledge in the Computocene. So, given the two somewhat antipodal preambles: "To posterity the arduous judgment" (*Ai posteri l'ardua sentenza*) [M1821], remains relevant to these developments. What feels like progress today may one day be seen as detour, or vice versa. There is a possibility that the positive and negative effects will balance each other out. In humility or perhaps in strategic deferral, let's hand over the verdict to history.

# From Acceleration to Attunement

> […] one of the most essential corrections to the character of humanity must be to strengthen the contemplative element.

Friedrich Nietzsche wrote this in *The Wanderer and His Shadow* (1880) [N1880]. Byung-Chul Han cites it at the end of *The Scent of Time* (2009) [H2009], as a kind of philosophical conclusion: a civilization that no longer knows how to be still is not moving forward, only spinning faster. At first glance, this might seem like the beginning of another essay lamenting our screens, our notifications, our fragmented attention. That is not exactly where I want to go. I want to try something else. I want to suggest, carefully and provisionally, that computers—which have long been blamed for our acceleration and distraction—might in the fullness of time help return us to something else entirely: a form of attunement, particularly in the context of discovery. Not by slowing the world down, but by taking over the parts of cognition that exhaust us, so that we might finally have space to notice what matters. Beyond that, by proposing new methods for acquiring knowledge where the roles of machines and humans working together remain to be determined. And right now, we are passing through an intermediate phase. How long it will take, I do not know. However, we are beginning to sense the potential of computers to participate in how we come to understand, not just by calculating but by shaping the very questions we ask and the ways we attend to the answers, towards new epistemic dynamics (here, *epistemic dynamics* refers to the patterns and processes by which knowledge, beliefs, and understanding evolve over time).

I realize this might sound "dangerously" close to positivism. *Positivism*, broadly speaking, is the belief that the human condition improves through the accumulation of scientific knowledge and the technologies that follow from it. It claims that what can be observed can be explained, and what can be explained can eventually be improved. In this view, progress is measurable. Problems are solvable. History moves forward, one innovation at a time. While there is nothing intrinsically wrong with positivism, in its more implicit forms it often drifts into over-optimism, a belief that technology alone, if allowed to advance far enough, will eventually fix or improve everything. Perhaps, to a certain extent, it does. That is not what I want to say. At least I hope not. If the reader finishes this essay with the impression that I believe computers will solve everything then I have failed to understand something, not just conceptually, but psychoanalytically. It would mean that I have not yet understood what I am trying to say, or why I needed to say it at all, or how to say it. Clearly, I write this not from a place of certainty, but from the middle of the noise.

In this evolving landscape, computers won't just be tools for automation or engines of statistical correlation. They are becoming something else: *scientific companions*. Machines to observe and discover. Not "simply" by crunching vast datasets or replacing closed-form equations with brute-force empiricism, but by elevating reasoning itself. In a way, by offering a new kind of patience, one rooted in composure and calm and measured interaction between humans and machines. In this sense above, computation becomes less a matter of acceleration/management and more a matter of *presence*. A presence that will extend beyond the screen, that accompanies us even when we're not interacting with it directly. Just as friendship persists in the silence between meetings, so too will these machines blur the boundary between online and offline, present and absent, calculation and contemplation. They will not replace us in discovery. They will sit beside us, quietly, while we learn to think together.

Just as the Anthropocene names an era shaped by human impact on the planet, the *Computocene* names an era shaped by the proliferation of computation, not only as massive infrastructure, working day and night in the background and gradually taken for granted, like aqueducts, but also as a mode of perception, both personal and collective. Computers will amplify our *powers of observation* in ways previously unimaginable. Once, we spoke of big data, of collecting everything with the hope that computer will find something useful in it. Soon, we will speak of machines that autonomously decide what is worth observing in the first place, asking us for an opinion and, perhaps, for permission to proceed based on what they have already seen. They will not only gather information but guide our attention. They will support our reasoning, lifting some of the cognitive weight that slows us down, freeing us to engage with structure, interpretation, wild guesses and meaning. This point offers a further look into a problem Herbert Simon identified half a century ago [S1971]. As he famously observed,

> What information consumes is rather obvious: it consumes the attention of its recipients. Hence a wealth of information creates a poverty of attention and a need to allocate that attention efficiently among the overabundance of information sources that might consume it.

Of course, all of this will not happen while we are idle. The rules of engagement will be different. These machines will not be mere typewriters, extensions of our memory and tools to transmit it to other human beings. They will aid reasoning not simply by spotting details or managing our thoughts more efficiently, but by subtly guiding our thoughts, suggesting directions we might not have otherwise considered. Innovation will be required in how such machines interact with us, and I am not even sure that mimicking human behavior is the right path forward—but that is another discussion entirely. Note that this not a functional shift aimed merely at boosting productivity. Paradoxically, it's through our interactions with machines that we become more aware to what feels mechanical, repetitive, or alienating—and, by contrast, what we value as distinctly human. Rather than passively receiving automation, we'll use it not just to save effort, but to clarify which efforts are meaningful.

(Speaking of effort and attention: a friend recently shared his enthusiasm for AI summarization. I appreciate summarization when I need to fix my bicycle or address an electrical issue at home, situations where I want (1) to find specific information quickly and (2) to receive clear, essential instructions. However, in scientific work, summarization isn't always beneficial because engaging with the thought process can be as valuable as understanding the final results.)

Said that, these notes are not a manifesto, nor a forecast. They are fragments from within this transition, and scattered reflections on what it will mean to live in a time when part of observation itself is automated and mediated. Importantly, the Computocene is not defined by the rise of *superintelligence*, again, term that we may or may not like. It will be (at least in part) defined by a new kind of attention, one in which human intelligence turns to the higher bits, to storytelling, which is something we are very good at, to policy making, and perhaps to the assignment of meaning. In this light, the Computocene appears not as an era of human obsolescence, but as a period of cognitive reorientation, where our distinctly human capacities for narrative understanding, ethical judgment, and meaning-making become more central, not less.

Here I want to focus specifically on science, also because when it comes to how (big) innovation begins, science often marks the first step. Execution, policy, and productization follow with their own importance. Science, however, enters at a different moment, early, when uncertainty is at its peak. It sets initial directions. It influences what becomes thinkable, what attracts attention, what gathers support. And because of how it works, it tends to build up momentum. Like a large ship, once it begins to turn, the course that follows becomes increasingly difficult to alter. This momentum makes early decisions in scientific pursuit particularly consequential. While scientific exploration may appear to meander randomly, it follows deeper patterns shaped by these initial choices. The seeming randomness of discovery masks an underlying directionality that, once established, becomes increasingly fixed. With the risk of sounding pedantic, this is why careful consideration at the outset is crucial: the apparent freedom to explore masks the profound influence of our initial directional choices.

[*Reader:* Your "Computocene" sounds like it's one chakra alignment away from digital enlightenment. Another one drifting into New Age territory.]

[*Author:* Perhaps you're right. But I'm writing this in April, in Seattle, where sunshine is more prophecy than weather. The rare appearance of the sun today might indeed be compromising my capacity for philosophical rigor.]

## Epochs of Attention

Let's suppose we're living in the Anthropocene. Everyone says we are, and there's something comforting in having a name for your era. It's like being told you're in Act III. The only question left is whether it's a tragedy or a comedy, and most likely, it's a comedy of errors. Not in the Shakespearean sense, with mistaken identities and happy reunions, but in the modern sense: a cascade of unintended consequences and half-written scripts.

The Anthropocene is defined by human impact on the atmosphere, the insects that vanish, the supply chain that stutters (which is a human system, but one operating at planetary scale), the debris in orbit, the millions of tons of carbon dioxide quietly dispersed into the air each year, the vast, slow-moving gyres of plastic drifting across the Pacific, stitched together by currents and neglect. It's an age of *acceleration*, confusion, and, unmistakably, *extraction*. More recently, it's become an age where we don't have time and where whatever "free time" remains is instantly filled with something else, usually more doing, which seems another form of extraction. And of course, it's an age of consequence.

Heidegger called it *Gestell* (enframing) a mode of revealing in which everything, including human beings, appears as *standing-reserve*: resources to be extracted, optimized, or held in readiness. In this view, the world is no longer encountered, but ordered; not experienced, but surveyed. Objects become inputs, and people become functions. "He himself will have to be taken as standing-reserve," Heidegger wrote in *The Question Concerning Technology* (1954) [H1977]. He warned that this shift does not begin with exploitation, but with perception (possibly, attention, to go back to Herbert Simon). What is dangerous is not the machine, but the worldview it carries with it. This is a quiet redefinition of value in terms of efficiency, prediction, and control. Under enframing, the possibility of seeing differently begins to recede (for example, a tree is no longer experienced as a tree, but as lumber). Reality arrives preprocessed and we lack the ability to read it differently.

Let's focus for a minute on the terminology. The term *Anthropocene* dates back to the early 2000s. Linguistically speaking, it's still a neologism. It was introduced by Paul Crutzen, a Dutch atmospheric chemist and Nobel laureate. The story goes that he interrupted a meeting where people were still talking about the Holocene, the stable, post-Ice Age epoch that began roughly 12,000 years ago. "No," he said, "we are in the Anthropocene." His point was simple: humans had become a geophysical force. We weren't just inhabiting the planet. We were editing its stratigraphy, not metaphorically, but literally. We were leaving behind a trace in the geological record: concrete, aluminum, ash, isotopes, plastic polymers that won't biodegrade, and the carbon signature of an atmosphere altered by combustion. These topics fall outside the scope of this essay, and we will leave them aside. The term first appeared in print in a short piece co-authored by Crutzen and Eugene Stoermer in 2000, in the *IGBP Newsletter* [CS2000]. Future layers of earth will carry our fingerprints, as if the planet had been marked by a kind of authorship.

Geologists have debated whether the Anthropocene deserves formal status: which layer of sediment counts, which spike in carbon or plastic or technofossil will be the one to mark it. But the cultural adoption of the term has never really depended on geological precision. It works because it names a feeling that we've tipped something. That the feedback loops are now global. That the world, as we knew it, is being managed by us. In this sense, the Anthropocene is not just a geological concept. *It's a psychological one.* A story we tell ourselves about having a lot of *agency*. Yet, even as the Anthropocene unfolds, another layer is forming. Not geological but informational. Something is shifting in how the world is sensed, measured, and made legible, not only to us, but to the machines we've built to observe it on our behalf as well. It's a kind of long process.

Thus, in the Computocene, knowledge production becomes tightly coupled to systems, where observation is shaped even more by the structure of our tools. It is not simply that data is abundant (and, in fact, data has a cost) but that models and machines end up to maximally condition what counts as signal. Measurement and observation mediated by machines becomes inseparable from prediction. So, Computocene, for me, here, is a psychological concept as well. The consequences are not limited to cognition; they propagate into infrastructure, energy consumption, and so on. Still, the central change that I aim to highlight is of epistemic nature: an adjustment in how knowledge is defined, validated, and operationalized within computational systems. In essence, I would like to capture a specific shift (self-quote):

> the reconfiguration of epistemic practices through computational mediation, which, if approached with care and critical awareness, does not merely automate thought but can reopen a space for *attunement*: a renewed sensitivity to patterns, feedback, and certain subtle structures of reality. In this light, computation is a tuning mechanism, capable perhaps of drawing our attention back to what matters, back to forms of understanding that resonate more naturally with our cognitive dispositions, instead of pulling us further away. When computation is allowed to reconfigure the *logic of inquiry* itself—not just its speed or scale—it becomes a medium for epistemic renewal, enabling us to discover ways of knowing that, potentially, had been obscured by usual scientific practices.

The word *attunement* once belonged to the realm of the intimate. It described a way of being sensitively aligned with the world. In philosophical terms, it was not a passive state but an active orientation, a condition of openness to what emerges. For Heidegger, *Stimmung* (attunement) revealed how we are always already situated, thrown into a world that discloses itself not in data but in feeling. In the Computocene, attunement acquires a new inflection. It becomes *mediated*. Not lost, but transformed. We are learning to attune ourselves not only to what is directly sensible, but to what becomes visible through machines.

These instruments do not merely extend our faculties; they reconfigure our access to reality. They filter, highlight, simplify, and sometimes amplify. They reduce noise, detect patterns, and extract something to which we can assign meaning—something that would otherwise remain imperceptible. In this sense, computation is not a barrier. It is an aid. It helps us re-enter the world rather than retreat from it. What we are attuning to is not the machine itself, but the world as revealed by the machine, through simulation, inference, and models. This is not a betrayal of perception. It is its continuation by other means. If this epoch is to have a human character, it may come from this very capacity: to remain attuned, through devices, to what matters. In this paradox lies the promise of the Computocene. Machines won't teach us to feel, but they might help us see what we otherwise could not, and, perhaps, to borrow Heidegger's language, ease some of the pressure of the enframing that technology has traditionally imposed.

[*Reader:* In your treatment above, you're suggesting that computers can help us reconnect with ourselves and enhance our presence. But isn't this just saying that if we use computers wisely, we can better focus on our tasks? Or perhaps you mean that with the right interface and human-computer interaction, we can focus better on what matters? Either way, it seems like basic principles rather than a novel insight.]

[*Author:* Possibly. Definitely, Joseph Licklider has been proven already right in saying [L1960],

> Man-computer symbiosis is an expected development in cooperative interaction between men and electronic computers in which the capabilities of the man and the computer are tightly coupled to solve problems much better than either man or the computer could do alone.]

[*Reader:* According to Wittgenstein's later philosophy [W2009], computers may never truly understand the physical world because understanding is not merely a matter of processing symbols or representing objects, but of participating in shared *forms of life*. For Wittgenstein, the meaning of a word arises from its use within specific social and practical contexts (what he called *language-games*) which are rooted in embodied human experience. We grasp what a "table" is not just by recognizing its shape, but by living in a world where we eat at tables, move them, trip over them, and share these interactions with others. Computers, by contrast, operate through syntactic manipulation or statistical inference, without embodied experience or access to the social practices that give our concepts meaning. They may simulate responses that appear meaningful, but they do not inhabit the world in the way humans do, and thus cannot genuinely play our language-games. In this view, understanding is not just computation (it is lived participation) and without that, machines remain outside the conditions that make meaning possible.]

[Author: I haven't claimed that computers will understand anything. They may not. What I've hinted to is their capacity to extend and augment our capabilities, regardless of whether they possess understanding in any meaningful sense.]

**After Observation**

Today, faced with complexity we can't hold in a single frame, we reach for proxies. We build layers of mediation not just to control the world, but to make it intelligible. The idea is no longer to experience the thing itself, but to manage its representation. This is not detachment. It is survival, somehow. The sheer scale of systems (for example, ecological, social, technical) exceeds our sensory and cognitive limits. In trying to keep up, we begin to mistake the structure for the substance. The simulation becomes the site of knowledge and it provides new opportunities for abstract and generalization.

By the way, our next breakthrough might come from learning to do more with less. Future discoveries may enable us to select data more judiciously, identifying crucial correlations that eliminate the need for exhaustive collection. Instead of drowning in data, we might develop more refined ways of choosing what to measure and observe. This "less is more" approach could mark a shift from pure accumulation to intelligent selection, where the quality of our questions matters more than the quantity of our data. In this way, the future of knowledge might not lie in gathering ever more information, but in becoming more astute about what information truly matters. Still, simulation will be central, in my opinion.

This movement toward simulation is not just conceptual. It is infrastructural. Nvidia CEO Jensen Huang has begun to describe modern data centers as "AI factories". These are not factories in the traditional sense of manufacturing physical goods, but facilities that generate simulations. These are the industrial sites of simulation: places where models are trained, refined, and deployed to interpret and, hopefully, predict the world.

[*Author:* I believe science is fundamentally about two things: describing what we observe and predicting what will happen. Explanation is a different and more difficult story.]

We now know that this process requires enormous amounts of data (so, physical machines) and energy. It starts to represent a *methodological inversion*: instead of forming a hypothesis and applying it outward, machines begin with everything and compress it inward into something that resembles interpretation. In the extreme expression of this, knowledge no longer emerges from a guess refined by testing, but from an accumulation refined by scale. The machines don't start with meaning. They arrive at it, often without needing to understand it, or even define what understanding is.

Recall that Norbert Wiener in The Human Use of Human Beings (1950) [W1950] warned that the danger was in our willingness to defer understanding to its outputs, rather than in the machine's power. His concern wasn't about the sophistication of computational systems, but about human tendency to surrender intellectual agency to them. The real threat, as he saw it, wasn't that machines would become too powerful, but that humans would become too passive in their relationship with technology.

So, returning to our main discourse, the trend in the future will move away from direct understanding of the world and toward its simulation. We will increasingly delegate knowledge production to machines through modeling. What once required interpretation now proceeds by optimization. What once depended on judgment now begins with inference. Observation no longer serves explanation; it initiates prediction, which is OK. *We will generate in order to predict*. We will need to be very careful, since this could be one of the biggest risks associated with the Computocene.

We rely on second-order observations that substitute for systems too complex to grasp firsthand, which is something we need to be careful about. This shift does not stem from laziness. It emerges from necessity and lack of verification. At this point, the systems we inhabit move too quickly and interlock too densely to be apprehended through direct experience. In genomic medicine, for instance, the models used to study disease risk are built to analyze population-wide patterns and possible pathways rather than to predict any individual's specific medical future. Researchers work with vast databases of genetic variations, observing correlations that emerge only through statistical aggregation. The attempts to understand biological reality take the form of a probability distribution. *Reality take the form of a probability distribution*. In financial modeling, for instance, the systems used to study market behavior are built to explore multiple possible scenarios and long-term trends rather than to predict next week's stock prices. Researchers work with ensembles of simulations, observing patterns that emerge through the aggregation of many variables and interactions. The attempts to understand market reality take the form of a probability distribution.

A tool of analysis are agent-based models, which traditionally begin with simple initial conditions and ask computers to generate possible scenarios through iteration. By systematically varying these initial conditions, we can explore multiple evolutionary paths of the system. This approach has gained wide acceptance across disciplines, from ecology to economics, where complex behaviors emerge from basic rules of interaction. The power lies not just in the computation itself, but in our ability to repeatedly test different starting points and observe how patterns emerge and stabilize over time. However, in the Computocene we may be entering a different paradigm of simulation. Instead of starting with simple, human-defined initial conditions, modern simulations may require vast amounts of data as their starting point, with initial conditions suggested by the computer itself. This represents a fundamental shift: from humans setting basic parameters and observing what emerges, to computers analyzing large datasets to determine the most relevant starting conditions.

Parenthetically, it is not difficult to imagine a future where even *data is no longer collected, but synthesized*. Not because we lack access to the world, but because it is faster and more efficient to invent plausible traces than to wait for real ones. In this imagined future, entire models are trained on data that no sensor ever recorded, that no person ever experienced. The result may still function; it may even outperform systems grounded in empirical input. The cost of this is subtle. We begin to model *simulations of simulations.* The system no longer observes reality; it refines an internally coherent fiction (see *A negative preamble* above). At some point, we forget what was real to begin with.

I like literary images of this fact. In *Invisible Cities* (1972), Calvino describes imagined places in such detail that they begin to feel as substantial as the world itself [C1972]. Borges' *Tlön, Uqbar, Orbis Tertius* (1940) talks about a fictional encyclopedia that describes an invented world so completely that it begins to appear in our own. First in fragments, then as artifacts, then in thought, and slowly, it starts to overwrite the world [B1940]. Something similar could happen here: synthetic data becomes not an exception, but the default. Clearly, *a machine trained on imagination is not doing science; it is producing fiction.* Moreover, it may generate narratives even more detached from reality than those created by humans. When we write fiction, we remain anchored—however loosely—to personal experience and our perception of reality. Rarely do we completely lose this connection, except in altered mental states. Machines, however, operate without these constraints. Their detachment is not artistic but structural.

These are well-known ideas, expressed across many recent contexts. It is difficult to credit a single thinker with the observation that modern society no longer relies on primary perception, but on the observation of observation. We increasingly operate in environments where meaning is not drawn from events themselves, but from how others anticipate and interpret those events. What we know is not the thing itself, but how it has already been seen, framed, and circulated. The world appears, more and more, as a reflection of its own internal readings. Our role is to monitor how the monitoring is going. As a result, we begin to interact more with models than with the world itself. It is easier. More measurable. More scalable. In time, the model becomes the primary object of study. We test it, train it, refine it. It is faster to simulate weather than to wait for it. More efficient to predict behaviors than to ask questions. More convenient to generate synthetic data than to collect real observations. The world becomes a backup.

This almost calls to mind Plato's cave, the ancient image of people watching shadows on a wall, mistaking them for reality, unaware that the true source lies behind them, outside their field of view. Except here, the shadows are hyper-detailed, and the details come not from reality but from imagination. Over time, we stop asking what casts them. A small consolation, perhaps, is that perception has always been mediated, with or without machines. What we call reality may never have been more than a well-rendered projection. It's funny to recall Heisenberg's remark,

> What we observe is not nature itself, but nature exposed to our method of questioning,

was about physics, not people.

Jean Baudrillard warned of this shift in *Simulacra and Simulation* (1981) [B1981]. He argued that simulation becomes dangerous when it no longer imitates reality, but replaces it. The sign of the real becomes more convincing than the real itself. A map as a territory. At some point, we begin to treat the model as the truth, and the world as an approximation. It is not difficult to imagine a distopic future, where scientific theories are no longer verified by direct observation, but by agreement between models. A theory is confirmed because it aligns with the output of another model, which itself was trained on historical patterns derived from earlier simulations. Observation gives way to recursive validation.

The risk is beyond epistemic collapse. The models still work. The outputs still arrive. However, the relationship between knowledge and reality begins to thin. The danger is to lose accuracy, and to stop noticing when the signal has drifted. Every new layer adds friction. Every observation of an observation increases the potential for distortion. Like in the children's game of 'telephone" (*passaparola* in Italian), where a message is whispered from person to person, information degrades with each transmission. Each participant retains and passes on what seems most significant to them, inevitably losing or distorting parts of the original message. The final version often bears little resemblance to what was first whispered. What matters is that the machine gives us answers at a lower economic cost. These risks suggest one of the philosophical cores of the Computocene:

> Not the rise of machine intelligence, but the redefinition of what it means to observe. Not the end of truth, but the beginning of different measures of fidelity and the inception of new forms of verification.

The Computocene will require us to develop techniques of empirical validation and reliable ways to verify truth (also) through computational methods.

## Wor(l)ds Matter

As the model becomes the lens through which we see the world, it's worth asking how we came to think of these systems in the first place. Before we can understand what machines are becoming, we might need to revisit what we believed they were. To understand how we arrived here, in an age defined by indirect observation and machine reasoning, it helps to look at the words we've used along the way.

Language preserves assumptions long after we stop noticing them. The names we give to machines reflect what we think they are, or what we once hoped they would do. So before going further, it's worth pausing over the term itself: computer. A word that still carries its arithmetic ancestry, even as its role is mutating into something else entirely. The word *computer* comes from the Latin *computare*, meaning to count, to sum up, to reckon. It once referred to people. In the 17th century, a computer was a person who performed calculations. It was quiet work: tables, mostly. In many languages, the root stayed close to arithmetic. *Ordinateur* in French, from *ordonner*, to order. *Rechner* in German, to calculate. *Calcolatore* in Italian. Each name gives something away. The German is blunt. The French, more aspirational. None of them, not even the English, captures what a computer has become. The ancient Greek is more honest, though probably accidental: *logismos*, reasoning, from *logos*—not just word, but principle, structure, a patterned way of seeing. In modern Chinese, the term is 计算机 (*jì suàn jī*), literally "calculation machine," pragmatic and mechanical. In Japanese, it's the same: 計算機 (keisanki), also "counting device." In modern Japanese, コンピュータ (konpyūta), which leaves little to imagination. Korean follows a similar logic. These are tools designed for structured mental work. The metaphors are stable, yet narrow.

In truth, none of these names fully account for what computation now does. A computer in the future I'm referring to is no longer just a calculator. It is an *epistemic machine*, in other words, a machine for producing conditions under which knowledge can appear. If we wanted to appeal to a different concept, we might call our era the *Logocene*, the age of structured reason, or perhaps of structured appearance. That term feels too abstract, and slightly theological. So, we settle for the *Computocene*, a hybrid of Latin and Greek, an etymological abomination. Still, it is closer to what people might recognize. It's something familiar. To the best of my knowledge, the term "Computocene" first appears in academic literature in the work of Comber and Eriksson in their paper *Computing as Ecocide* [CE2013]. They use the term in a hypothetical context, stating:

> No-one wants to find that in ten or twenty or a hundred years that we look back and rename this phase of earth's existence from Anthropocene to Computocene in recognition of, for instance, the genotoxic legacy of the accumulation of e-waste.

While Comber and Eriksson use the term to highlight the potential long-term environmental consequences of computing, particularly the accumulation of e-waste, my use differs substantially, and perhaps even mistakenly. I use the term to refer to the impact of computation on human cognition, behavior, and our ways of interacting with the world, while recognizing that this perspective is necessarily partial. I also like tracing the term back to something ethimologically earlier than *computare*, which is *calculus*, Latin for "pebble". A stone used for counting. This is a reminder that computing begins with the material. From this follows the idea that computers are physical machines, embedded in the world, and should be understood as instruments for observing that very world—an observation that lies at the heart of our notin of attunement.

## Epistemic Machines

The shift from the computer as a mere calculator to an epistemic machine mirrors a profound change in our strategies for handling complexity. Why the delegation to machines? The defining feature of the current intersection of the Computocene and the Anthropocene is cognitive overload. Consequently, we increasingly rely on machines and here, I'm referring specifically to core computational devices, distinct from their more derivative forms like laptops and cell phones, which often function as everyday utensils and, ironically, contribute to a different kind and more well-known cognitive overload.

Bruno Latour, for his part, warned that modernity's real rupture was not science itself, but the belief that we could stand outside the world we study—see *We Have Never Been Modern* [L1991]. In the Anthropocene, the species with the most agency may also be the one most estranged from its own effects. The Computocene is what happens when computers become our primary means of making sense of the world. When seeing the world means simulating it. In the Anthropocene, we shape the planet. In the Computocene, we model it, perhaps ahead of experience. In working with simulations, we will need to decide how closely we want to remain tethered to reality. This decision defines a continuum: on one end, models that serve as heuristic tools—useful for generating suggestions, structure, or inspiration; on the other, simulated experiments that produce data meant to be interpreted, confirming or challenging some hypotheses. Along this spectrum, the physical hardware of machines begins to matter more. This is where quantum computing, as specialized hardware, becomes relevant:

> Quantum hardware is primarily an instrument for scientific experimentation with quantum phenomena.

I will stress this point, especially to the idea that quantum computers, in the fulness of time, may be better understood not as algorithmic engines with immediate relevance to human affairs, but as devices for exploring certain selective features of the natural world.

[*Reader:* Remind me. What is a quantum computer?]

[*Author:* Here is a definition. A *quantum computer* is a machine that modifies the global state of an ensemble of objects (natural or artificial) whose behavior is governed by the laws of quantum physics, such as atoms or photons. It does so through specialized hardware that encodes information directly in the quantum states of objects in the ensemble, which are then manipulated through sequences of physical operations applied to small subsets of those objects.]

[Reader: Are quantum computers more intelligent than non-quantum computers?]

[Author: There are specific outputs that, given an input, a quantum computer can compute using fewer elementary operations, potentially in less time. This doesn't imply intelligence, which, to be honest, I'm not even sure what it is.]

You see that this is very far from superintelligence, a term that is difficult to define and often better suited to science fiction, with its narratives of machines rendering humans obsolete. (Unless with the functional definition "machines capable of reasoning, learning, and adapting across any domain," proposed by Dario Amodei in a 2024 interview with Lex Fridman [AF2024]). We already live alongside machines that extend our capabilities. Bicycles make us faster. Pocket calculators perform arithmetic we struggle to do in our heads. Computers, from the beginning, have allowed us to operate beyond our native limits. It's the "any domain" part that matters. Personally, I have no issue with this. I accept it, and even expect it. But I believe the Computocene's opportunity should be interpreted differently. In this context, quantum computers embody the shift—less toward intelligence, more toward orientation and observation. As Harvard physicist Mikhail Lukin puts it, "computers will take us to places of the universe we cannot go without," not in a physical sense, clearly, but referring to epistemic exploration (where I removed the term "quantum" from the original quote) [L2025]. Whether computers, in general, will be more or less intelligent than we are, is terra incognita. The idea forces us to look in a very specific direction and this now become an absolutely pervasive narrative. One simpler thing that matters ("simpler" because at least it doesn't require us to go down into the weeds of the definition of intelligence) is where computers allow us to look and in what form they allow us to acquire knowledge. Machines can expand our intellectual horizons, challenge our existing cognitive frameworks, and potentially unlock new modes of human understanding and creativity, while, maybe, will computationally make easier for us a number of problems.

## Computers in the Wild

Computers are physical objects. They are governed by the laws of physics. This is not incidental, it is foundational. If we want to understand the world, one path is to build machines that reflect those same laws. Quantum computers are a firm step in that direction.

[*Reader:* Isn't forcing quantum computers in this story?]

[*Author:* While quantum computing is the area I've spent the most time considering, I believe it fits naturally here. Lukin's insight about exploring new corners of the universe captures exactly what I mean about the broader relationship between humans and computing technologies and quantum computers seem to me a plausible example.]

They do not just simulate physical processes; they *function* as physical processes. They model the world not from the outside, but by inhabiting its logic. In this sense, *all buildable* computers are not abstract engines running code, but material devices engaged in (scientific) experimentation. This idea has been described on and on, with different sauces, to the point that it became folklore.

In the early 1980s, Richard Feynman and Yuri Manin independently proposed that certain quantum physical systems could not be simulated efficiently by classical computers, and thus a new kind of machine, built from the principles of quantum mechanics, would be needed (see [F1982]). Then, Paul Benioff, Charles Bennett, and Rolf Landauer, amond others, further shaped the field, arguing that computation itself must obey thermodynamic constraints. Landauer, in particular, insisted that "information is physical," challenging the notion of abstract, disembodied logic [La1991]. These early voices set the stage for quantum computing as a discipline concerned not with speed, but *with the alignment of computation and nature*. Other thinkers, such as those associated with the Santa Fe Institute, *e.g.*, Chris Langton, Stuart Kauffman, *et al.*, have similarly explored the idea that computation is a natural phenomenon. Soap films solving minimal surface problems are often cited as analog examples.

In each case, the computation is not metaphorical or representational; it is embedded in matter. *Computation becomes a way to* see *the world by building physical systems that express its structure.* (For a book combining insights from physics and computation see [MM2011].)

Paradoxically, then, the computer brings us back to the material (recall that *calculus* means stone). It reconnects us with physical laws. It invites a closer look at the substrate of knowledge. Whether we call these machines intelligent, or compare them to us, is beside the point. What matters is where they allow us to look. Now, the conceptualization of computers as tools for observation represents a fascinating thread in philosophical discourse that crosses multiple domains including epistemology, philosophy of mind, philosophy of science, and philosophy of technology. Rather than being a concept championed by a single philosopher, this perspective has emerged through various philosophical traditions and approaches. Said that I am quite certain that any attempt on my part to review the literature would fall short. At the very least, I should mention a traditional divide in Western philosophy, if only to begin ordering my thoughts. Bernard Stiegler notes that "at the beginning of its history, philosophy separates tekhnē from epistēmē, a distinction that had not yet been made in Homeric times" [S1994]. This foundational division positioned *technical know-how* (tekhnē) as something distinct from and often subordinate to *theoretical knowledge* (epistēmē).

The separation created a philosophical context in which tools and techniques were not considered proper subjects of epistemological inquiry. Problems of "how to" were traditionally excluded from the epistemic realm, they were considered matters of *craft* rather than *knowledge*. The conceptualization of computers as observation tools represents a kind of shift in this traditional division, bringing the technical back into the realm of knowledge-creation. Interestingly, algorithms embody "how to" in its purest form, since they are sequence of instructions. In passing, programming has even been described as a form of art by Donald Knuth, though his position was somewhat ambiguous. On the one hand, he titled his monumental work *The Art of Computer Programming*, framing programming as art, tautologically. On the other hand, he once remarked that "science is what we understand well enough to explain to a computer. Art is everything else [K1968].

One of the most explicit articulations of computers as tools for observation comes from Harold Abelson and Gerald Jay Sussman. In their textbook *Structure and Interpretation of Computer Programs* [AS1985] they introduce the concept of *procedural epistemology*, which they define as "the study of the structure of knowledge from an imperative point of view." Abelson and Sussman argue that computation provides a fundamentally different way of knowing than traditional mathematical approaches. While "Mathematics provides a framework for dealing precisely with notions of *what is*," they suggest that "Computation provides a framework for dealing precisely with notions of *how to*."

This positions computation not just as a technique or application of knowledge, but as a distinct mode of observation and knowledge-creation. Their concept of procedural epistemology suggests that by implementing processes computationally, we gain specific insights that would be unavailable through mere theoretical analysis. The computer becomes in this way an instrument for observing how processes unfold, revealing patterns and potentially constraints that might not be visible through other means of investigation.

This perspective invites several interpretations. One is the well-known line attributed to Feynman: *what I cannot build, I do not understand* [F1988]. Is there a better way than a computer to build something? (But I guess we need to come up with a satisfactory definition of "building".) Fine. But more than that, it points to our growing ability to construct *in silico* models of the world—which may be designed not in the abstract, but in direct relation to the physical substrate and internal structure of the machine we are using. The logic of the computation is not separable from the material properties of the hardware, seen as an instrument for observation. Consider, for example, dedicated chips for matrix multiplication; their architecture constraints how we think and ultimately observe. Indeed, a number of computer scientists have long sensed that computation is more than calculation. At its core, it reflects a deeper intuition: computers are tools for exploring the space of the possible, given constraints of syntactic and physical nature.

For example, in the pedagogical context, Seymour Papert described the computer as an "object to think with," a device that encourages a scientific state of mind by enabling concepts to be shaped through direct interaction and construction [P1980]. It's a nice take.

Jim Gray proposed what he called the "Fourth Paradigm" of science: *data-intensive discovery*. In his view, the vast amounts of data now generated by scientific instruments, simulations, and sensors demand a new methodology, where the ability to analyze data becomes central to how we produce knowledge. Computation, in this paradigm, is beyond a way to simulate the world, but the only observational instrument we know to extract meaning from the overwhelming scale of empirical information [T2009]. The fourth paradigm differs fundamentally from simulation in several respects. While simulation begins with theoretical models to generate predictions, data-intensive discovery starts with vast amounts of real-world data from which insights are extracted. Simulation typically uses data to construct and validate models, whereas in the fourth paradigm, data itself drives the discovery process (but the boundaries are blurred). This approach integrates theory, experiment, and simulation with large-scale data analysis, forming a unifying framework that strengthens all scientific methods. It engages with more diverse and voluminous datasets than traditional simulation, demanding not only computational power but also advanced tools for data management and analysis (for example, the instance of a simulation problem could have a small input). Unlike simulation, which is often designed to test specific hypotheses, data-intensive science can do both, test hypotheses and uncover patterns or relationships that were not anticipated.

Rather than replacing simulation, it complements and extends it, opening new pathways for integrating and analyzing knowledge across disciplines, and navigate complexity at larger scales. A well-known example of the fourth paradigm in action is the discovery of exoplanets using data from NASA's Kepler space telescope. Kepler collected light curves from over 150,000 stars, generating a dataset too vast for traditional analysis. Researchers applied machine learning algorithms to detect subtle dips in brightness (signatures of planetary transits). This led to the identification of thousands of exoplanets, including Kepler-90i, part of the first known eight-planet system outside our own [SV2018]. Crucially, this discovery was not driven by a specific hypothesis, but emerged from pattern recognition across massive observational data, demonstrating how data-intensive methods can reveal what theory and simulation alone might miss.

Let us return to quantum computers. They embody, perhaps more than any other technology (that comes to my mind), the idea of computation as a means of observation. Quantum computing is often introduced as a promise to speed up certain tasks—a fact that, while important, is almost incidental. I love this quote from Michael Nielsen's *In What Sense Is Quantum Computing a Science?* (2018) [N2018]:

> […] quantum computing will be in considerable part a design science. That is, it'll be about discovering new types of objects and behaviour. This is a point of view that is perhaps unusual, even idiosyncratic. It will take many decades to tell if I am correct. But I believe it's a stimulating point of view, and likely to be correct.

The word *discovering* is used with care. It implies that these objects and behaviors are already there—latent, but inaccessible through classical means, obscured by complexity that overwhelms traditional hardware or lack of tools for observation. This marks another shift: *from computation as execution to computation as epistemology*. It's a powerful mental model.

David Deutsch described quantum computing as "a better way to understand the laws of physics" [D1997]. His argument (again) was not about speed, but about alignment: an ontological fit between the structure of the world and the structure of our machines. The consequence is not just technological. It is philosophical. Simulating a molecule on a quantum computer is not merely an exercise in numerical modeling. It is the enactment of a physical model. A portion of the world, therefore its dynamics and its logic, are instantiated in another physical substrate. The result is not a representation, but a second-order instantiation: a translation of nature into nature, matter into matter. Even if the second occurrence of the term "nature" refers to something engineering buy human beings. More strongly, as Scott Aaronson (among others) kept suggesting for about twenty years by now, the pursuit of quantum computing is not solely about achieving computational speedups over classical machines; it's also (and primarily) about probing the fundamental principles of the universe that allow such quantum phenomena to exist. It's worth noting what Tony Leggett wrote in the final chapter of his book *The Problems of Physics* (1987) [L1987]:

> I am personally convinced that the problem of making a consistent and philosophically acceptable 'join' between the quantum formalism which has been so spectacularly successful at the atomic and subatomic level and the 'realistic' classical concepts we employ in everyday life can have no solution within the current framework; that it is only a matter of time before the increasing sophistication of the technology available makes this not just a philosophically, but an experimentally, urgent question; that the resulting conceptual impasse will eventually generate a quite new description of the world, whose nature we cannot at present even begin to guess.

With about forty years having passed since Leggett's quote, the question arises: will quantum computers become the resolutive tools needed to address this, even if it takes another forty years? (Recall that Leggett received the Nobel Prize in Physics in 2003 for his pioneering work on superfluidity in helium-3.)

**A Metaphor**

In *The Aleph* [B1970], Borges tells the story of a man who visits the home of Beatriz Viterbo, a woman he once loved. After her death, he continues to visit each year on her birthday. There, he meets her cousin, Carlos Argentino Daneri, a second-rate poet with an ambitious plan: to describe every place on Earth in verse. Daneri claims he can do this because of something hidden in the cellar. This is a point in space called the Aleph. It contains all other points. From it, you can see everything that exists, from every angle, all at once. The narrator is skeptical but agrees to look.

In the cellar, he sees the Aleph. In a flash, he is overwhelmed by a vision of the entire universe: every object, every person, every event, simultaneous and infinite. The experience is indescribable. He leaves unsure whether Daneri understands what he has, or whether he himself will be able to write again after seeing so much.

The Aleph is a reasonable, though perhaps too indulgent, metaphor for the computer in the Computocene. Like the Aleph, the computer promises access—not to knowledge itself but to a new kind of visibility. It offers a way of seeing that reshapes what it means to observe, to describe, and perhaps even to know. This is not a conclusion, but a point of departure.